\documentclass[12pt, twocolumn, twocolappendix, numberedappendix]{openjournal}
\usepackage{graphicx,float}
\usepackage{xcolor}
\usepackage{hyperref}
\usepackage{upgreek}
\hypersetup{
    colorlinks=true,
    linkcolor=blue,
    filecolor=blue,      
    urlcolor=blue,
    citecolor=blue,
}
\usepackage{color,colortbl}
\DeclareGraphicsExtensions{.bmp,.png,.jpg,.pdf}
\usepackage{orcidlink}
\usepackage{amsmath}
\usepackage{framed}
\usepackage{ifthen,amsmath,amssymb, bm}

\newcommand{\vk} {\bm{k}}
\newcommand{\wt}[1]{\widetilde{#1}}

\newcommand{\D} {\mathcal{D}}

\newcommand{\vb} { {\bm{b}} }

\newcommand{\vtheta} {\bm{\theta}}

\DeclareRobustCommand{\Fig}[1]{Fig.~\ref{fig:#1}}

\DeclareRobustCommand{\Eq}[1]{Eq.~\ref{eq:#1}}

\definecolor{c2}{RGB}{0,168,50} 

\begin{document}
\title{Towards precision astrometry of scattered images of compact radio sources: scintillometry theory and prospects}

\author{Dylan L. Jow \orcidlink{0000-0003-3236-8769}}
\affiliation{Kavli Institute for Particle Astrophysics \& Cosmology, P.O. Box 2450, Stanford University, Stanford, CA 94305, USA}
\email{dylanjow@stanford.edu}
\author{Delon Shen \orcidlink{0000-0003-3963-4038}}
\affiliation{Department of Physics, Stanford University, 382 Via Pueblo Mall, Stanford, CA 94305, USA}
\affiliation{Kavli Institute for Particle Astrophysics \& Cosmology, P.O. Box 2450, Stanford University, Stanford, CA 94305, USA}
\affiliation{SLAC National Accelerator Laboratory, 2575 Sand Hill Road, Menlo Park, CA 94025, USA}
\email{delon@stanford.edu}
\begin{abstract}
Compact radio sources such as pulsars and FRBs undergo scintillation in the interstellar medium (ISM) when scattered images interfere at the observer. ``Scintillometry'' refers to the range of techniques to extract astrometric information---such as the angular positions of the images and distances to the scattering screen and source---from scintillation observations. Pulsar scintillometry has proven to be a powerful technique, revealing rich and unexpected scattering phenomenology in the ISM and also shedding light on the emission physics of pulsars. FRB scintillometry stands to be a similarly powerful probe of FRB emission, as well as structure on tiny scales in ionized media beyond our galaxy, such as the circumgalactic medium (CGM). However, nascent FRB scintillation studies are far from the sophisticated lensing geometry reconstructions that have been performed for scintillating pulsars. In this paper, we introduce a novel theoretical framework for scintillometry, demonstrating that the full astrometric content of scintillation observations is contained within a single underlying observable: the instantaneous spatial wavefield. We relate the instantaneous spatial wavefield to more familiar concepts from the pulsar scintillometry literature, such as the dynamic spectrum. Using this framework, we discuss prospects and limitations for FRB scintillometry, towards the goal of full astrometric reconstructions of FRB lensing geometries. We show how key degeneracies in two-screen scattering measurements can be ameliorated. In addition, we discuss the possibility of inferring dispersion measure gradients across scintillation screens, which may shed light on the highly unconstrained physics of the cool CGM phase on tiny ($\sim 100\,{\rm au}$) scales. 
\end{abstract}
\maketitle

\section{Introduction}
\label{sec:intro}

Interstellar scintillation is the frequency and position dependent intensity modulation that compact radio sources undergo due to the interference of multiple images formed by scattering in the ionized interstellar medium (ISM). Recently, observations of interstellar scintillation of FRBs have been used to place constraints on the size of the emission region of the source
\citep{Nimmo2025}. Scintillation has also been used to place constraints on the distance from the source to the scattering material that is responsible for the characteristic angular broadening and scattering tails observed in FRBs \citep{Sammons2023}. These constraints rely on the basic physical principle that interference effects are suppressed when the angular extent of the source is effectively resolved by the scattering screen. In particular, \citet{Nimmo2025} and \citet{Sammons2023} rely on two-screen arguments, where the source is said to only be able to scintillate if the scattering screen further from the source does not resolve the angular broadening due to the screen closer to the source. However, there are important limitations to these arguments. Firstly, anisotropic screens (i.e. screens that only scatter light in one direction) only have resolving power in one direction. Thus, two anisotropic screens that are perpendicular to each other on the plane of the sky act entirely independently. Secondly, plasma lenses that produce a small number of images can induce a second observed scintillation scale without decohering (i.e. suppressing the scintillated modulations), even when the lens is fully resolved by the scintillation screen. Finally, one cannot infer the resolving power of a scattering screen based solely on its scintillation bandwidth (i.e. the characteristic scale of the intensity modulations in frequency) alone---one must also measure the distance to the screen. In current two-screen scintillation studies, the distances are unknown, and one must assume that one of the screens resides within the Milky Way ISM to get constraints. In contrast to FRB scintillation studies, the technique of \textit{scintillometry} has been used in observations of pulsars to obtain full lensing geometry reconstructions \citep{Liu2016lensgeometry, HRZhu2023, YHChen2025}. Not only are distances to multiple screens measured, but also the angular positions of the individual scattered images on the sky are obtained. This information has been used to resolve pulsar emission regions \citep{Main2021crabemission}. Moreover, examples of two-screen scattering in the ISM have been found where both screens fully resolve each other while maintaining scintillation, problematizing two-screen FRB scintillation analyses \citep{HRZhu2023}. 

Ideally, we would like to be able to leverage pulsar scintillometry techniques to perform the same lensing geometry reconstruction for FRBs. Not only is this important for resolving key uncertainties in two-screen constraints on the source emission region, it has important implications for FRB scattering studies. FRB scattering may be an important probe of the cool ($T \sim 10^4\,{\rm K}$) gas component of the circumgalactic medium (CGM), as it will be sensitive to fluctuations on tiny scales (potentially down to sub-au scales) \citep{vedantham2019radio, jow2024refractive, ocker2025microphysics}. However, there has yet to be an unambiguous detection of FRB scattering in the CGM, despite the many absorption line studies which indicate the ubiquitous presence of dense, ionized structures in the CGM on relevant scales \citep{Rigby2002, Hennawi2006, BowenCheclouch2011, tumlinson2017circumgalactic, mccourt2018}. It has been argued that this gas may not significantly scatter FRBs \citep{ocker2022radio, masribas2025refractive, ocker2025microphysics}; however, even if these structures do not produce the characteristic exponential scattering tails that can be readily identified as scattering, they may still produce more subtle plasma lensing signatures. For example, a smooth plasma lens may produce only a few discrete images, as opposed to the large number of images needed for scattering to manifest as an exponential scattering tail. If the burst is also scattered significantly by plasma in the Milky Way ISM and the FRB's host environment, the plasma lens would be difficult to identify without a full lensing geometry reconstruction. Thus, scintillometry techniques have the potential to reveal rich plasma lensing phenomenology, not only in the source environment and the Milky Way ISM, but also in the CGM of intervening halos. Even if CGM halos never produce any scattering or plasma lensing effects, \citet{Baker2025dmgradient} show that resolving the scattered images from the a scintillation screen in the ISM can be used to measure changes in the dispersion measure across au-separations. Applying this technique to FRBs would enable constraints on the turbulent fluctuations of the CGM at these tiny scales. Leveraging scintillation to resolve intervening lensing systems may also be a powerful probe of cosmology, enabling direct distance measurements to microlensed FRBs \citep{Tsai2024microlensing}.

Towards this goal of multi-screen FRB lensing geometry reconstructions, we introduce a novel theoretical framework for understanding scintillometry. Specifically, we introduce the instantaneous spatial wavefield. We show that, for a single screen, a measurement of the instantaneous spatial wavefield can be used to directly infer the angular positions of the scattered images in addition to the screen distance. We relate the instantaneous spatial wavefield to the so-called dynamic spectrum---an observable well-known in the pulsar scintillometry literature---by demonstrating that the dynamic spectrum is a lower-dimensional version of the instantaneous spatial wavefield. In doing so, we clarify the assumptions necessary for pulsar scintillometry to be effective, which further enables forecasts for our ability to reproduce these techniques with FRBs. 

The paper is organized as follows. In Section~\ref{sec:scint}, we describe the basic theoretical framework for scintillometry. We explain the different scintillometry techniques, followed by a discussion of the feasibility of these techniques in FRB applications. In Section~\ref{sec:sims}, we perform simulations of FRB scintillometry with more realistic assumptions about the burst properties. In Section~\ref{sec:CGM_dispersion}, we discuss the prospects for measuring gradients in the dispersion measure of FRBs across scintillation screens (similar to the measurement performed for pulsars by \citet{Baker2025dmgradient}), and we predict the contributions of the ISM, CGM, and the intergalactic medium (IGM) to the differential dispersion measure. 

\section{Scintillometry theory}
\label{sec:scint}

Scintillometry---a portmanteau of ``scintillation" and ``astrometry"---refers to a set of techniques that may ultimately be used to determine the positions of the scattered images of compact radio sources due to interstellar scintillation. Phase-coherent images of the FRB or pulsar form via scattering through an ionized medium and arrive at the observer where they interfere. In the same way that regular radio interferometry extracts high-precision astrometric measurements by interfering observations at multiple baselines, scintillometry leverages the interference pattern of images formed by an interstellar scattering screen to determine the positions of those images and, in some cases, the screen and source distances. However, in contrast to radio interferometry, where different pairs of antennas are manually selected to interfere, in scintillation, all of the scattered images interfere at the observer, simultaneously. This leads to degeneracies which must be broken by measuring the interference pattern as a function of position. In this section, we will describe the theory of scintillometry and how astrometric information is encoded in the observed interference pattern.

First, we introduce the instantaneous spatial wavefield, $V(\nu, {\bm b})$, which is the complex wavefield (containing both amplitude and phase) of the incident radiation at an observer with a location ${\bm b}$ in the observer plane and frequency $\nu$. The \textit{instantaneous} spatial wavefield should be contrasted with the \textit{dynamic} wavefield, which is often referred to in the interstellar scintillation literature, which we will discuss later. As we will show, the instantaneous spatial wavefield contains all of the astrometric information and is the underlying observable that all scintillation observations seek to measure. 

\begin{figure}
    \centering
    \includegraphics[width=\linewidth]{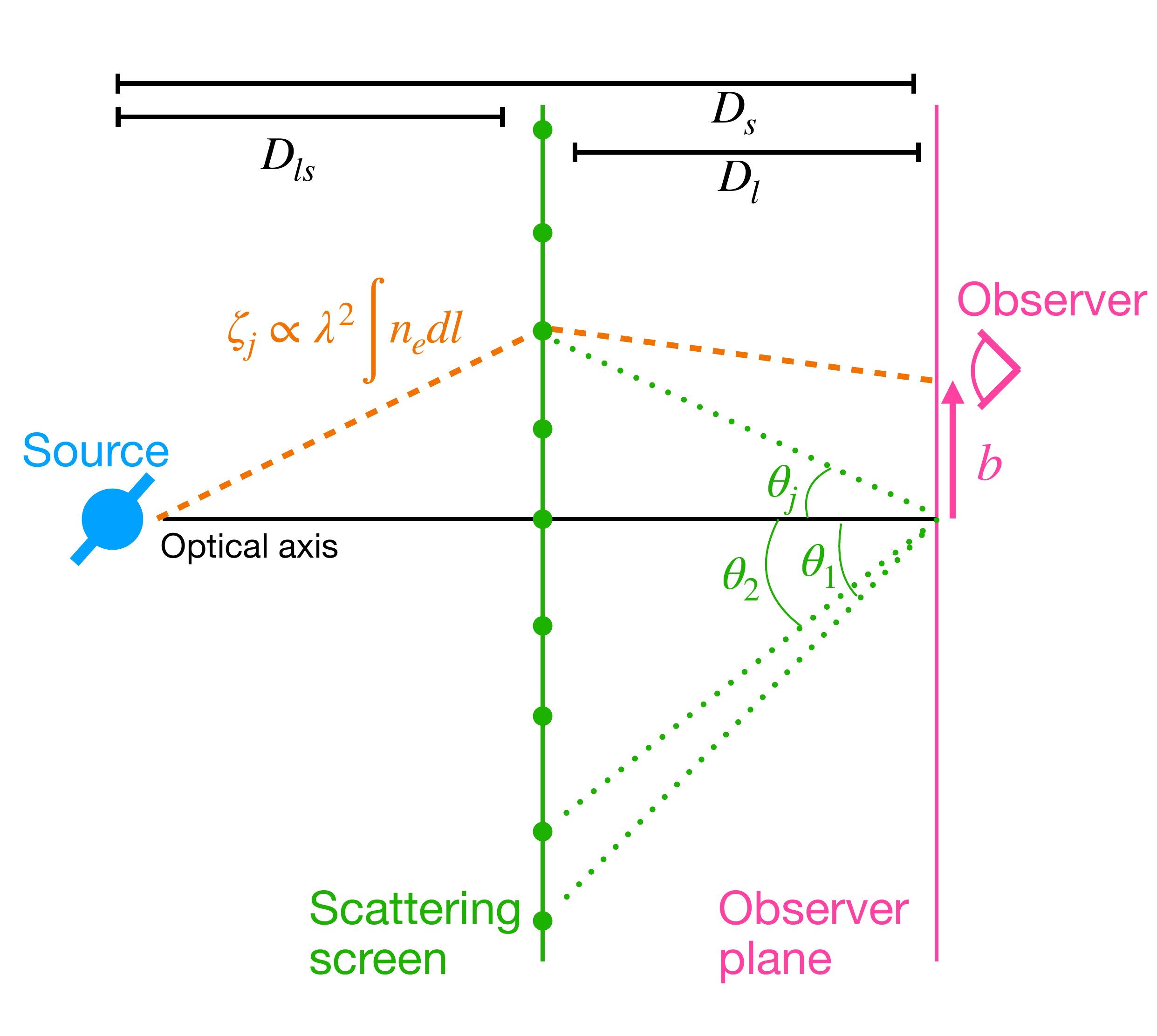}
    \caption{Diagram of the lensing geometry through a scattering screen. Fixed images (green circles) are formed on the scattering screen at angles $\theta_j$ defined relative to the optical axis. Eq.~\ref{eq:phase} gives the phase of rays arriving at the observer passing through image $j$ relative to the ray arriving at the observer plane through the optical axis. The phase has three components: 1. the quadratic $\theta_j^2$ term due to the images' offset from the optical axis, 2. the ${\bm b} \cdot {\bm \theta_j}$ term due to the observer's offset, ${\bm b}$, from the optical axis, and 3. the accumulated phase delay $\zeta_j$ due to the different dispersion measure (${\rm DM} = \int n_e dl$) along the each ray. The distances $D_l$, $D_s$, and $D_{ls}$ refer to the distances from the observer plane to the scattering screen, the source, and the distance between screen and source, respectively. They appear in Eq.~\ref{eq:phase} through the \textit{effective} distance, $D = D_l D_{s}/D_{ls}$. For sources at infinity, $D \approx D_l$.}
    \label{fig:scattering_diagram}
\end{figure}

Now, assuming that the incident electric field is a coherent sum of a finite number of scattered images with intensities $\mu_j$ and phases $\Phi_j$, then the spatial wavefield is given by
\begin{equation}
    V(\nu, {\bm b}) = \sum_j \mu^{1/2}_j e^{i \Phi_j(\nu, {\bm b})},
    \label{eq:ISW_sum}
\end{equation}
where the image phases are a function of observer position and frequency \citep{T:2025pew}:
\begin{equation}
    \Phi_j(\nu, {\bm b}) = 2 \pi\nu \left[ \frac{D}{2 c}{\bm \theta}^2_j  + \zeta_j - \frac{1}{c}{\bm b} \cdot {\bm \theta}_j \right].
    \label{eq:phase}
\end{equation}
Fig.~\ref{fig:scattering_diagram} is a diagram showing the relevant terms that result in a phase delay between the images. The first term is the geometric, $\theta^2_j$ term, which is the relative time delay between an image at $\theta_j$ and the central image at $\theta=0$. The second term is the dispersive delay, $\zeta_j$, which encodes the fact that each image propagates through a different total electron column as it travels from source to observer, resulting in a differential dispersion measure. This dispersive delay is itself frequency dependent, with $\zeta_j \propto \nu^{-2}$. The final term is the position of the observer relative to the optical axis, which results in an additional delay proportional to ${\bm b} \cdot {\bm \theta}_j$. 

Note that in writing the spatial wavefield in this way we have made a few assumptions. Firstly, we have assumed that the bursts have a flat intrinsic spectrum (i.e. are essentially delta functions in the time domain), so that variations in the observed spectrum only arise due to the interference of the images (Eq.~\ref{eq:ISW_sum}). For pedagogical clarity, we will maintain this assumption throughout this section; however, we will discuss the more general case in Section~\ref{sec:sims}. Secondly, we have assumed that the scattering comprises a finite number of images due to a single screen at a fixed distance, $D_l$. Pulsar scintillation observations support the conclusion that, at least for some sight-lines, interstellar scintillation is indeed dominated by a single scattering screen along the line of sight \citep{Brisken2010, Stinebring2022survey, YHChen2025}. Finally, we have assumed that the scattered images are at a fixed location, $\theta_j$, and do not change with the observer position or observing frequency. This is a generic expectation of refractive lensing where the scattered images are effectively anchored to the geometric position of the lenses \citep{Pen:2013njl} and is consistent with observations \citep{Baker2023thetatheta, YHChen2025}.

Now, we can expand $\Phi_j$ to linear order in $\nu$ and ${\bm b}$ as:
\begin{align}
    \Phi_j &\approx \Phi^0_{j} + 2 \pi \tau_j (\nu - \nu_0) + 2 \pi {\bm k}_j \cdot ({\bm b} - {\bm b_0}), \label{eq:phase_expansion}\\
    \tau_j &\equiv \frac{1}{2\pi} \frac{\partial \Phi_j}{\partial \nu} \Big|_{{(\bm b}_0, \nu_0)} = \frac{D}{2c}{\bm \theta}_j^2 - \zeta_j - \frac{1}{c} {\bm b}_0 \cdot \theta_j, \label{eq:tau_group}\\
    {\bm k}_j &\equiv \frac{1}{2\pi} \nabla_{\bm b} \Phi_j \Big|_{{(\bm b}_0, \nu_0)} = -\frac{1}{\lambda_0} {\bm \theta}_j, \\
    \Phi_j^0 &= 2 \pi \nu_0 \tau_j.
\end{align}
We have defined the time delay, $\tau_j$, which is the total \textit{group} delay with which each image arrives relative to the central image. Note that because the dispersive delay is inversely proportional to the frequency squared, $\zeta_j \propto \nu^{-2}$, it appears with a negative sign in the group delay. We have also defined $k_j$, which is simply the angular position of each image divided by the central wavelength of the observation, $\lambda_0 = c\nu_0^{-1}$. It follows that
\begin{equation}
    V(\nu, {\bm b}) \approx \sum_j \mu_j^{1/2} e^{i \Phi_j^0} e^{i 2 \pi \tau_j (\nu - \nu_0)} e^{i 2 \pi {\bm k} \cdot ({\bm b} - {\bm b_0})}.
    \label{eq:sw_expanded}
\end{equation}
From this, it may already be clear that by measuring the instantaneous spatial wavefield, we can measure both the time delays of the images and their angular positions. The time delays set the scale of the oscillations of the instantaneous spatial wavefield in the frequency domain, while the angular positions set the scale of the oscillations in the spatial domain. This is easiest to see in Fourier space.

We define the conjugate spatial wavefield (CSW) as the 2D-Fourier transform of the instantaneous spatial wavefield: $\tilde{V}(\tau, {\bm k}) \equiv \mathcal{F}_\nu[\mathcal{F}_{\bm b} [V]]$. From Eq.~\ref{eq:sw_expanded}, it follows that 
\begin{equation}
    \tilde{V}(\tau, {\bm k}) = \sum_j \mu_j^{1/2} e^{i \Phi_j^0} \delta(\tau - \tau_j) \delta^{(2)}({\bm k} - {\bm k}_j),
    \label{eq:CSW}
\end{equation}
where $(\tau, {\bm k})$ are the conjugate variables to $(\nu, {\bm b})$.  Thus, each image forms a delta function in the 3D conjugate space at $(\tau_j, {\bm k}_j)$. Now, setting ${\bm b}_0 = 0$ and assuming, for the moment, that $\zeta_j = 0$, one finds that
\begin{equation}
    \tau_j = \frac{\lambda^2_0 D}{2 c} {\bm k}^2_j.
    \label{eq:tau_v_sigma}
\end{equation}
That is, the geometric delay is simply quadratic in ${\bm k}$. Thus, by measuring $\tilde{V}$ one can obtain the image locations ${\bm \theta}_j = \lambda_0{\bm k}_j$, and one can obtain the distance parameter, $D$, by measuring the curvature of the paraboloid formed by the images in $(\tau, {\bm k})$-space (``delay/k-space"). Assuming the dispersive delays are small relative to the geometric delays, each $\zeta_j$ can be recovered by measuring the deviation of the observed time delays, $\tau_j$, from the inferred paraboloid.  

\begin{figure*}
    \centering
    \includegraphics[width=\linewidth]{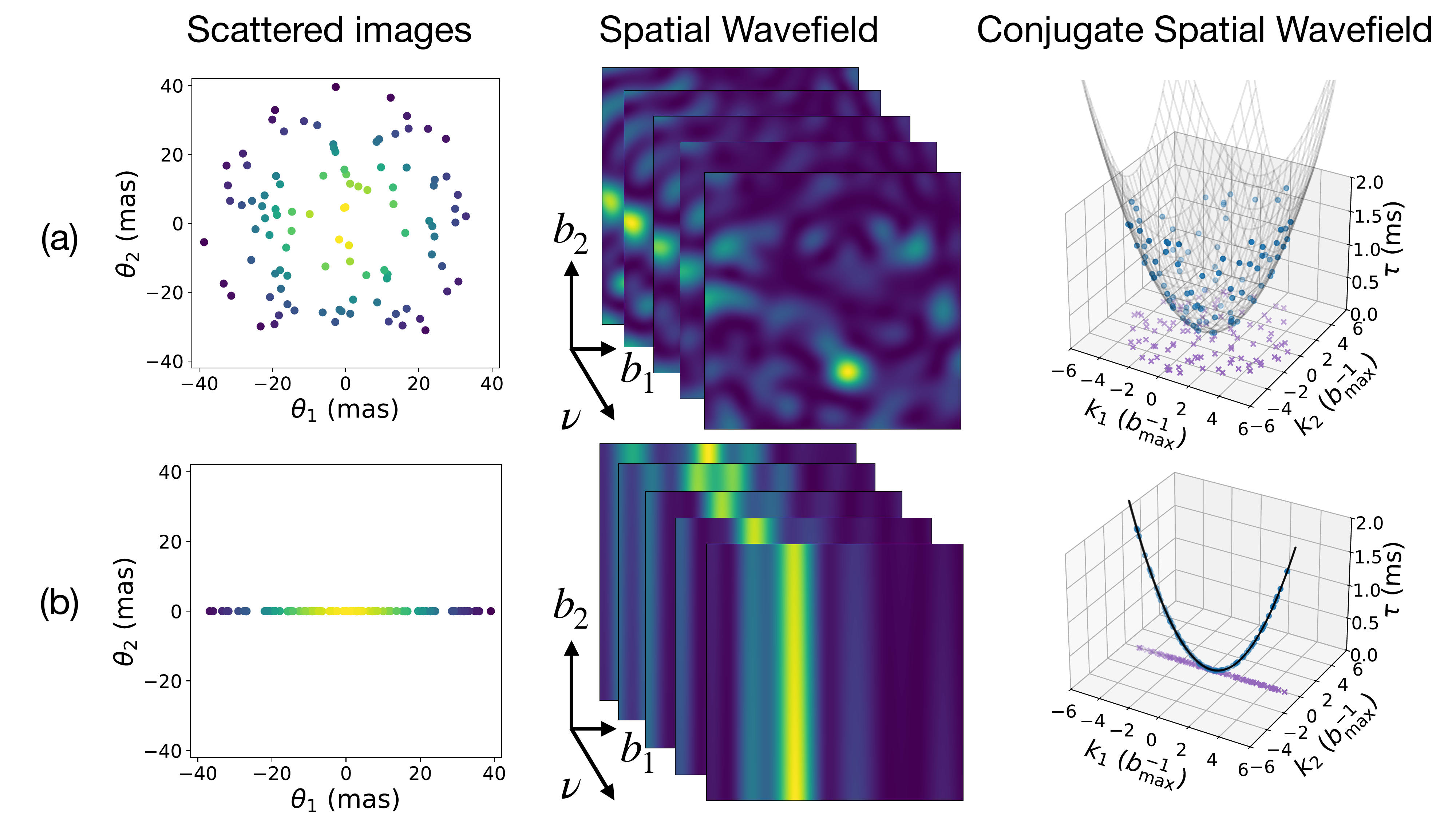}
    \caption{The instantaneous spatial wave field and its conjugate for an isotropic scattering disc (top row) and an anisotropic scattering disc (bottom row). The left panel shows the angular position of the scattered images on the sky, with the colour indicating their magnification $\mu_j$. A radial Gaussian profile is chosen for the magnifications with a width of $\theta_{\rm sc} = 20\,{\rm mas}$, and $N_{\rm im} = 100$ images are uniformly distributed in a disc of radius $2 \theta_{\rm sc}$ (in the isotropic case) and a line of width $4 \theta_{\rm sc}$ (in the anisotropic case). The middle panel shows the instantaneous spatial wavefield that arises from these scattered images, computed via Eq.~\ref{eq:ISW_sum}, where we choose $D = 1\,{\rm kpc}$, $\lambda = 75\,{\rm cm}$, and $\zeta_j = 0$. We compute the spatial wavefield over a grid of observer positions with dimension $10^4 \,{\rm km}\times 10^4\,{\rm km}$. We plot the spatial wavefield for different values of $\nu$ to show its evolution in frequency. The right panel shows the points of non-zero power (blue) in the resulting conjugate wavefield (Eq.~\ref{eq:CSW}). The purple points show the blue points projected onto the ${\bm k}$-plane and the mesh paraboloid shows the quadratic relationship between the delay $\tau_j$ and ${\bm k}_j$ (Eq.~\ref{eq:tau_v_sigma}).}
    \label{fig:kfnu-space-diagram}
\end{figure*}

Fig.~\ref{fig:kfnu-space-diagram} shows two examples of how scattered images manifest in the instantaneous spatial wavefield and the conjugate spatial wavefield. The top row shows an isotropic scattering screen, where the scattered images are randomly populated in a disc of $40\,{\rm mas}$ on the sky. Taking the effective distance $D = 1\,{\rm kpc}$ and the central observing frequency $\nu_0 = 400\,{\rm MHz}$, one can compute the instantaneous spatial wavefield (i.e. the interference pattern as a function of observer position and frequency), as well as the conjugate spatial wavefield. The position of the non-zero points in the conjugate spatial wavefield in the ${\bm k}$-plane are simply the angular positions of the images on the sky scaled by $-\lambda_0$. The time delay is then given by a paraboloid with curvature $\lambda^2_0 D/c$. The second row shows the same phenomenon, except the scattering disc is squeezed by a factor of 100 in one direction. The result is that the previously 3D spatial wavefield becomes essentially two-dimensional, with only the frequency axis and one spatial axis being relevant. The effect in the conjugate space is to flatten the paraboloid by the same factor of 100. 

The resolution with which one is able to measure the image time delays and angular positions, therefore, boils down to simple Fourier analysis. The timing resolution is set by the inverse bandwidth of the observation, $\delta \tau \sim \Delta \nu^{-1}$. This, in turn, is limited by the intrinsic bandwidth of the radio pulse, but also, in practice, may be further limited by our assumption that the time delay is effectively independent of frequency over the observing bandwidth in expanding the phase as in Eq.~\ref{eq:phase_expansion}. The angular resolution is set by the largest baselines in the array: $\delta \theta \sim \lambda / b_{\rm max}$. At $f = 400\,{\rm MHz}$, one attains $\sim\,{\rm mas}$ angular resolution with baselines $>10^5\,{\rm km}$. Thus, it will be challenging to resolve individual scattered images using an instantaneous measurement of the spatial wavefield with terrestrial baselines. Nevertheless, distinguishing between the isotropic and anisotropic scattering scenarios shown in Fig.~\ref{fig:kfnu-space-diagram} remains plausible.

\subsection{Sub-${\rm mas}$ resolution with pulsar scintillometry}
\label{sec:pulsars}

While terrestrial baselines will limit the angular resolution that can be achieved by an instantaneous observation of the spatial wavefield, scintillometry techniques have been developed that have been used to resolve the scattered images to sub-${\rm mas}$ resolution for some scintillating pulsars \cite[see e.g.][]{Brisken2010, Simard2019scint, YHChen2025}. In this section, we will describe these techniques and how they relate to the instantaneous spatial wavefield. Essentially, the relative motion of the Earth, scatterer, and pulsar system enables one to sample the spatial wavefield over time with longer baselines than would be achievable through an instantaneous, terrestrial measurement. Rather than measuring a snapshot of the spatial wavefield with an array of radio receivers, one measures the spectrum as a function of time at a single receiver, as it sweeps through the spatial interference pattern. This, of course, relies on the fact that pulsars repeat, so that the scintillation pattern can be measured over time. The techniques of pulsar scintillometry are discussed extensively elsewhere in the literature \citep{Walker2004scinttheory, Cordes2006arcs, Brisken2010, Sprenger2021thetatheta}; however, we give a high-level overview here in order to illustrate some key conceptual points that will be useful when we consider the case of FRB scintillometry. 

\begin{figure*}
    \centering
    \includegraphics[width=0.7\linewidth]{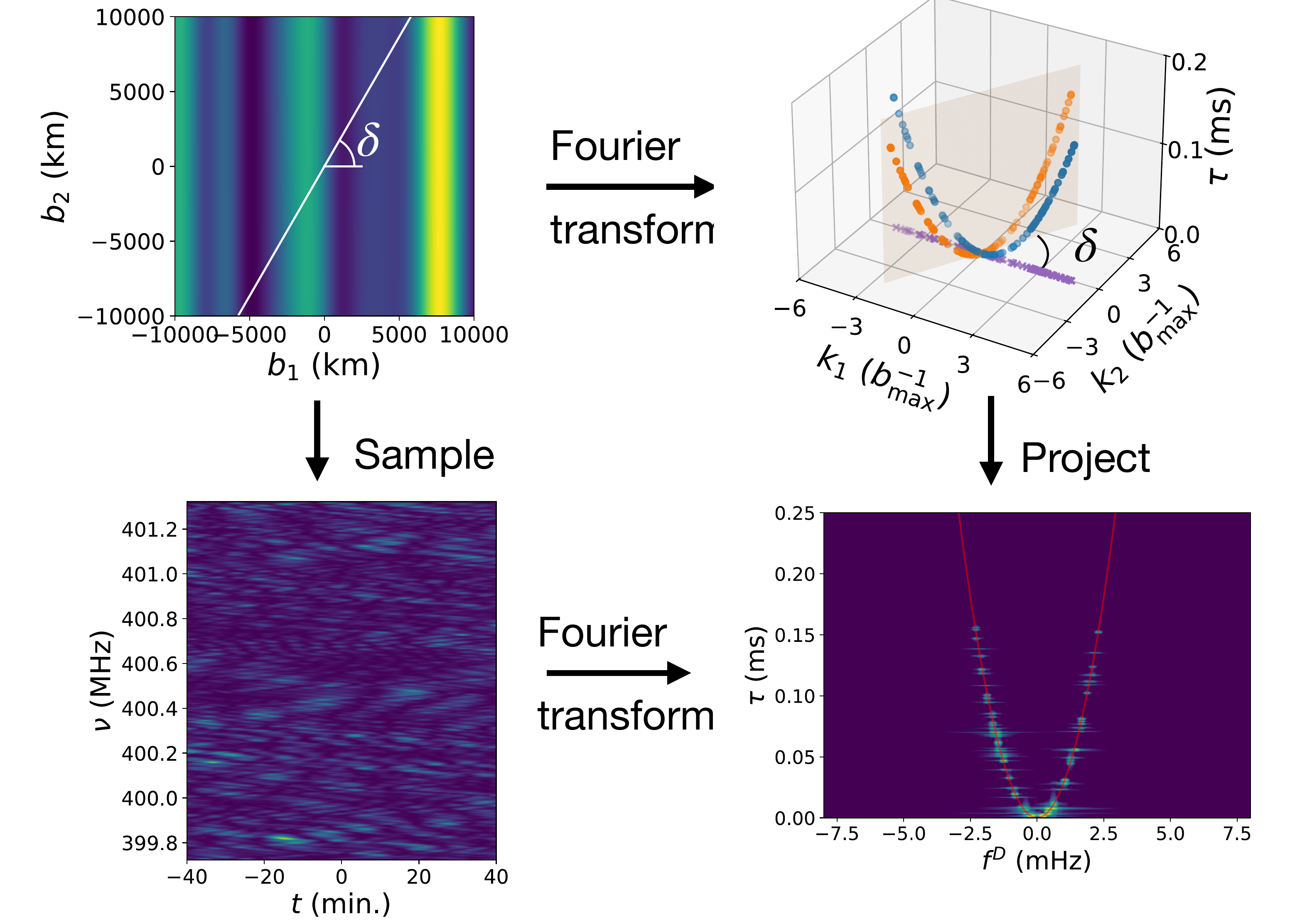}
    \caption{Diagram showing the relationship between the instantaneous spatial wavefield, the conjugate spatial wavefield, the dynamic wavefield, and the conjugate dynamic wavefield. (Top left) The 3D instantaneous spatial wavefield shown as a function of observer position ${\bm b}$ for a single frequency $\nu$ corresponding to $\lambda = 75\,{\rm cm}$. The scattering is due to an anisotropic line of $N_{\rm im} = 100$ images with a length $\theta_{\rm sc} = 20\,{\rm mas}$ and an effective distance $D = 100\,{\rm pc}$. (Top right) The non-zero points of the full conjugate spatial wavefield (blue). The purple points show the blue points projected onto the ${\bm k}$ plane, and the orange points show the blue points projected onto a vertical plane at an angle $\delta$ from the scattering direction. (Bottom left) The 2D dynamic wavefield that would be observed for an effective velocity of $V_{\rm eff} = 20\,{\rm km / s}$ at an angle $\delta$ from the scattering direction. That is, the dynamic wavefield is obtained by sampling the spatial wavefield along the white trajectory in the top left panel. (Bottom right) The conjugate dynamic wavefield obtained by taking the 2D Fourier transform of the bottom left panel. The conjugate dynamic wavefield is related to the conjugate spatial wavefield by taking the projected orange points in the top right panel and re-scaling the horizontal axis by $V_{\rm eff}$.}
    \label{fig:dynamic_spectrum_example}
\end{figure*}

Pulsar scintillometry seeks to make astrometric inferences from the \textit{dynamic} wavefield, $V(\nu,t)$, which is the spectrum of the pulsar seen by a single observer as a function of time. The dynamic wavefield can be related to the spatial wavefield via:
\begin{equation}
    V(\nu, t) = \sum_j \mu_j^{1/2} e^{i \Phi_j({\bm b}(t), \nu)}.
\end{equation}
Comparing this to Eq.~\ref{eq:ISW_sum}, we see that the dynamic wavefield is simply the spatial wavefield, except we have replaced the two-dimensional co-ordinate ${\bm b}$ with a single time co-ordinate by re-writing the position ${\bm b}$ as a function of time:
\begin{align}
    {\bm b}(t) &= {\bm V}_{\rm eff} (t - t_0), \\
    {\bm V}_{\rm eff} &= {\bm V}_\oplus + \frac{D_l}{D_{ls}} {\bm V}_{\rm s}  - \frac{D_s}{D_{ls}} {\bm V}_l,
    \label{eq:Veff}
\end{align}
where $V_{\rm s}, V_\oplus, V_l$ are the velocities of the source, Earth, and screen, respectively. We assume that over the course of the observation, the relative motion of all three systems can be described by a single, constant velocity. We also assume that the spatial wavefield is itself unchanged over the duration of the observation. The first column of Fig.~\ref{fig:dynamic_spectrum_example} shows the relationship between the spatial wavefield and the dynamic wavefield for an anisotropic scatterer. The top panel shows the spatial wavefield in ${\bm b}$-coordinates (at a single frequency), and the white line shows the trajectory of the effective trajectory ${\bm b}(t)$. The bottom panel shows the dynamic wavefield, which is simply the top panel (the spatial wavefield) sampled along the white curve across different frequencies. 

Similarly to before, we can define the conjugate dynamic wavefield to be the 2D Fourier transform of the dynamic wavefield:
\begin{equation}
    \tilde{V}(\tau,f^D) = \sum_j \mu_j^{1/2} e^{i \Phi^0_j} \delta(\tau - \tau_j) \delta(f_t - f^D_j),
\end{equation}
where $f^D$ is the conjugate variable to time and is referred to as the Doppler shift. Each individual image attains a Doppler shift:
\begin{equation}
    f^D_j \equiv \frac{1}{2\pi} \frac{\partial \Phi_j}{\partial t} =\frac{1}{\lambda_0} {\bm V}_{\rm eff} \cdot {\bm \theta}_j.
\end{equation}
The time delay is
\begin{equation}
    \tau_j = \frac{D}{2c} {\bm \theta}_j^2 + \zeta_j - \frac{t_0}{c} {\bm V}_{\rm eff} \cdot {\bm \theta}_j.
\end{equation}
Since we replaced two spatial dimensions with one time dimension, it is no longer possible to measure ${\bm \theta}_j$ directly from the dynamic wavefield. Instead, one is only able to measure the scalar ${\bm V}_{\rm eff} \cdot {\bm \theta}_j$, where $V_{\rm eff}$ is generally unknown. Likewise, the geometric part of $\tau_j$ is, in general, not quadratic in $f^D_j$. Thus, it may seem at first glance that the dynamic wavefield has basically none of the useful information that the spatial wavefield had. Although the dynamic wavefield is a sampling of the spatial wavefield, it is a sampling via an \textit{a priori} unknown trajectory. However, as Fig.~\ref{fig:dynamic_spectrum_example} illustrates, useful information can be extracted in the case that the scattering is highly anisotropic. 

If the scattering is one-dimensional, so that ${\bm \theta}_j = \theta_j \hat{n}$, then (setting $t_0 = 0$ and $\zeta_j = 0$):
\begin{align}
    \tau_j &= \frac{\lambda_0^2 D}{2 V^2_{\rm eff, \parallel} c} (f^D_j)^2, \\
    f^D_j &= \frac{V_{\rm eff, \parallel}}{\lambda_0} \theta_j,
\end{align}
where $V_{\rm eff, \parallel} \equiv {\bm V}_{\rm eff} \cdot \hat{n}$ is the component of the effective velocity in the direction of the scattering screen. Thus, for one-dimensional scattering, the power in the conjugate dynamic wavefield lies along a parabola in $(\tau, f^D)$-space, and measuring the curvature of that parabola gives a measurement of $D V^{-2}_{\rm eff, \parallel}$---the effective distance is now degenerate with the effective velocity. The location of the images along the $f^D$ axis gives a measurement of their angular position, multiplied by $V_{\rm eff, \parallel}$. An alternative way of viewing this is illustrated in the top right panel of Fig.~\ref{fig:dynamic_spectrum_example}. The blue paraboloid comprises the scattered images in $(\tau, {\bm k})$-space that one would find if one was able to measure the full spatial wavefield. If, instead, one is only able to measure the spatial wavefield along a linear trajectory at some angle $\delta$ to the scattering direction (shown in white in the top left panel), then the conjugate dynamic wavefield is given by projecting the conjugate spatial wavefield in the 3D $(\tau, {\bm k})$-space onto a 2D vertical plane at an angle $\delta$ to the scattering direction (the orange surface). The horizontal axis of this plane is then scaled by $V_{\rm eff}$. The result is a parabola with the same curvature as the initial $(\tau, {\bm k})$-space curvature, scaled by $(V_{\rm eff} \cos \delta)^{-2} = V^{-2}_{\rm eff, \parallel}$. From this, it can be easily seen why observing a parabolic arc in the conjugate dynamic wavefield is indicative of anisotropic scattering: if the scattering were isotropic (i.e. forming an isotropic paraboloid in $(\tau, {\bm k})$-space), then its projection onto $(\tau, f^D)$-space would not be a thin parabola. Nevertheless, even in the full isotropic case, the envelope of the power would be parabolic. 

The advantage of measuring the dynamic wavefield over time, as opposed to instantaneously sampling the spatial wavefield with a terrestrial array, is that one is able to leverage the motion of the Earth to achieve much larger effective baselines. For an effective velocity of $V_{\rm eff} \approx 20\,{\rm km\,s^{-1}}$, an hour-long observation yields an effective maximum baseline of $72,000\,{\rm km}$, corresponding to milliarcsecond angular resolution. However, by measuring the dynamic wavefield, one introduces a degeneracy between ${\bm \theta}_j$ or the effective distance $D$ and the effective velocity ${\bm V}_{\rm eff}$. However, in some cases, the velocity of the pulsar is well-measured, allowing $V_{\rm eff}$ to be inferred and the degeneracy lifted. One must also know the distance to the pulsar, as the \textit{effective} velocity depends on the distance between the scattering screen to the source (Eq.~\ref{eq:Veff}). If, however, the pulsar distance and velocity are known, then the full lensing geometry, including the distance to the scattering screen $D_l$, can be inferred from the dynamic wavefield, either by assuming the screen velocity is small (i.e. assuming the third term in Eq.~\ref{eq:Veff} vanishes) \citep{YHChen2025}, or by modeling the arc curvature's variation over several epochs \citep{Liu2016lensgeometry, HRZhu2023}.

One critical point that we have neglected in this section so far for the sake of clarity is that, in practice, one does not measure the incident wavefield, $V$, but rather the intensity, $I = |V|^2$, thereby losing the complex phase information. This is because pulsars undergo significant intrinsic variation in brightness and phase between bursts. It is only the average pulse intensity profile that is stable, and so, to measure the effect of interstellar scintillation on the variability, one must average over bursts in intensity. Thus, in practice, one actually measures the ``dynamic spectrum" (the square of the dynamic wavefield), and, subsequently, the 2D Fourier transform of the dynamic spectrum, referred to as the  ``secondary spectrum". By basic properties of Fourier transforms, the secondary spectrum is simply a convolution of the conjugate dynamic wavefield with itself: $\tilde{V} \star \tilde{V}$. For 1D scattering screens, this self-convolution leads to the classic inverted arclets that are seen in pulsar scintillation secondary spectra \citep[e.g.][]{Brisken2010, Stinebring2022survey}. However, techniques have been developed to recover the lost phase information and re-construct the dynamic wavefield itself, such as the $\theta-\theta$ transform \citep{Sprenger2021thetatheta, Baker2023thetatheta}, or cyclic spectroscopy \citep{Demorest2011cycspec, Turner2023cycspec}. The former relies on the assumption that the scattering is highly anisotropic, leveraging the simple quadratic relationship between the scattering angle and time delay. The latter relies on the stable periodic nature of pulsar signals to retrieve the phase without any additional assumptions on the nature of the scattering.

\subsubsection{Multi-baseline pulsar scintillometry}
\label{sec:multibaseline-psr-scint}

As we have seen, if the distance to the pulsar and its velocity are known, then measuring the dynamic wavefield is sufficient to reconstruct the lensing geometry. However, most pulsar distances and velocities are poorly known. Indeed, scintillometry may be regarded as a technique for \textit{obtaining} unknown pulsar distances. However, this cannot be done through observations of the dynamic wavefield at a single station. Instead, one must combine dynamic wavefield observations across multiple (terrestrial) baselines. To illustrate this technique, we will define the dynamic wavefield as measured by an observer at ${\bm l}$ as
\begin{equation}
    V(t,\nu;{\bm l}) = \sum_j \mu_j^{1/2} e^{i \Phi_j({\bm l} + {\bm V}_{\rm eff}(t-t_0), \nu)}.
\end{equation}
We will also define the cross-dynamic spectrum as
\begin{equation}
    W(\nu, t; {\bm l}) = V\left(\nu, t; -\frac{{\bm l}}{2}\right) V^*\left(\nu, t; \frac{{\bm l}}{2}\right).
    \label{eq:cross-secondary-spectrum}
\end{equation}
Note that the uncertain intrinsic phase of the pulse cancels in the cross-dynamic spectrum. The cross-dynamic spectrum was first introduced by \citet{Brisken2010}. Here, we are relating it to the underlying spatial wavefield by writing the dynamic wavefield at ${\bm l}$ as a particular sampling of the former over a linear trajectory with some constant offset ${\bm l}$

Now, the Fourier transform of the cross-dynamic spectrum (the cross-secondary spectrum) is then
\begin{align}
\begin{split}
    \tilde{W}(\tau, f^D; {\bm l}) = \sum_{j,k} &\mu_j^{1/2} \mu_k^{1/2} \delta(\tau - \tau_{jk}) \delta(f^D - f^D_{jk}) \\
    &\times \exp \left\{ i \left[ \Phi^0_j(-\frac{\bm l}{2}) - \Phi^0_k(\frac{\bm l}{2}) \right] \right\},
\end{split}
\label{eq:Wftfv}
\end{align}
where 
\begin{align}
\nonumber
\tau_{jk} &\equiv \tau_j - \tau_k\\
\nonumber
&= \frac{D}{2c}( {\bm \theta_j}^2 - {\bm \theta_k}^2) - (\zeta_j - \Delta_k )+ \frac{1}{2c} {\bm l} \cdot ({\bm \theta_j} + {\bm \theta_k}) \\
&\approx \frac{D}{2c}( {\bm \theta_j}^2 - {\bm \theta_k}^2) - (\zeta_j - \Delta_k)
\label{eq:taujk}
\\
f^D_{jk} &\equiv f^D_j - f^D_k = \frac{1}{\lambda_0} {\bm V}_{\rm eff} \cdot ({\bm \theta}_j - {\bm \theta_k}).
\label{eq:fDjk}
\end{align}
We drop the ${\bm l}$ term in $\tau_{jk}$ as it will generally be much smaller than either the geometric or dispersive delays. For a baseline of $l = 10,000\,{\rm km}$ and a maximum scattering angle of $10\,{\rm mas}$, this delay is $< 1\,{\rm ns}$. However, the complex phase of the cross-secondary spectrum is given by
\begin{align}
    \Phi_{jk}({\bm l}) &\equiv  \Phi^0_j\left(-\frac{\bm l}{2}\right) - \Phi^0_k\left(\frac{\bm l}{2}\right) \\
    &= 2 \pi \nu_0 \left[ \frac{D}{2c} ({\bm \theta}^2_j - {\bm \theta}^2_k) - (\zeta_j - \Delta_k) + \frac{1}{2c} {\bm l} \cdot ({\bm \theta}_j + {\bm \theta_k}) \right].
\label{eq:Phi0jk}
\end{align}
Here, we keep the ${\bm l}$ term. While it leads to a total delay that is negligible compared to the other terms, it can nevertheless lead to a measurable difference in phase. In other words, even nanosecond delays are important when determining phases for waves with nanosecond periods.

Now, Eq.~\ref{eq:Wftfv} is a sum over pairs of images, $({\bm \theta}_j, {\bm \theta}_k)$, that satisfy Eqs.~\ref{eq:taujk}, \ref{eq:fDjk}. For highly anisotropic scattering (${\bm \theta}_j = \theta_j \hat{n}$) there is only one pair of images that produces a given $(\tau_{jk}, f^D_{jk})$. This is a critical assumption that enables one to directly infer the image locations from the cross-secondary spectrum. One must simply measure the symmetric part of this phase,
\begin{equation}
    \psi_{jk} = \Phi_{jk} + \Phi_{kj} = \frac{2 \pi}{\lambda_0} {\bm l} \cdot ({\bm \theta}_j + {\bm \theta}_k),
\end{equation}
which is achieved by computing
\begin{equation}
    \psi({\bm l}) = {\rm arg}\left[C(\tau, f^D;{\bm l})\right],
    \label{eq:symmetric-phase}
\end{equation}
where 
\begin{equation}
    C(\tau, f^D; {\bm l}) = \tilde{W}(\tau, f^D; {\bm l}) \tilde{W}(-\tau, -f^D; {\bm l}).
\end{equation}
Note that since the transformation $\tau_{jk} \to -\tau_{\rm jk}$ and $f^D_{jk} \to -f^D_{jk}$ is the same as swapping images $j \to k$, the phase of $C$ is precisely the symmetric part of $\Phi_{jk}$. 

\begin{figure}
    \centering
    \includegraphics[width=\linewidth]{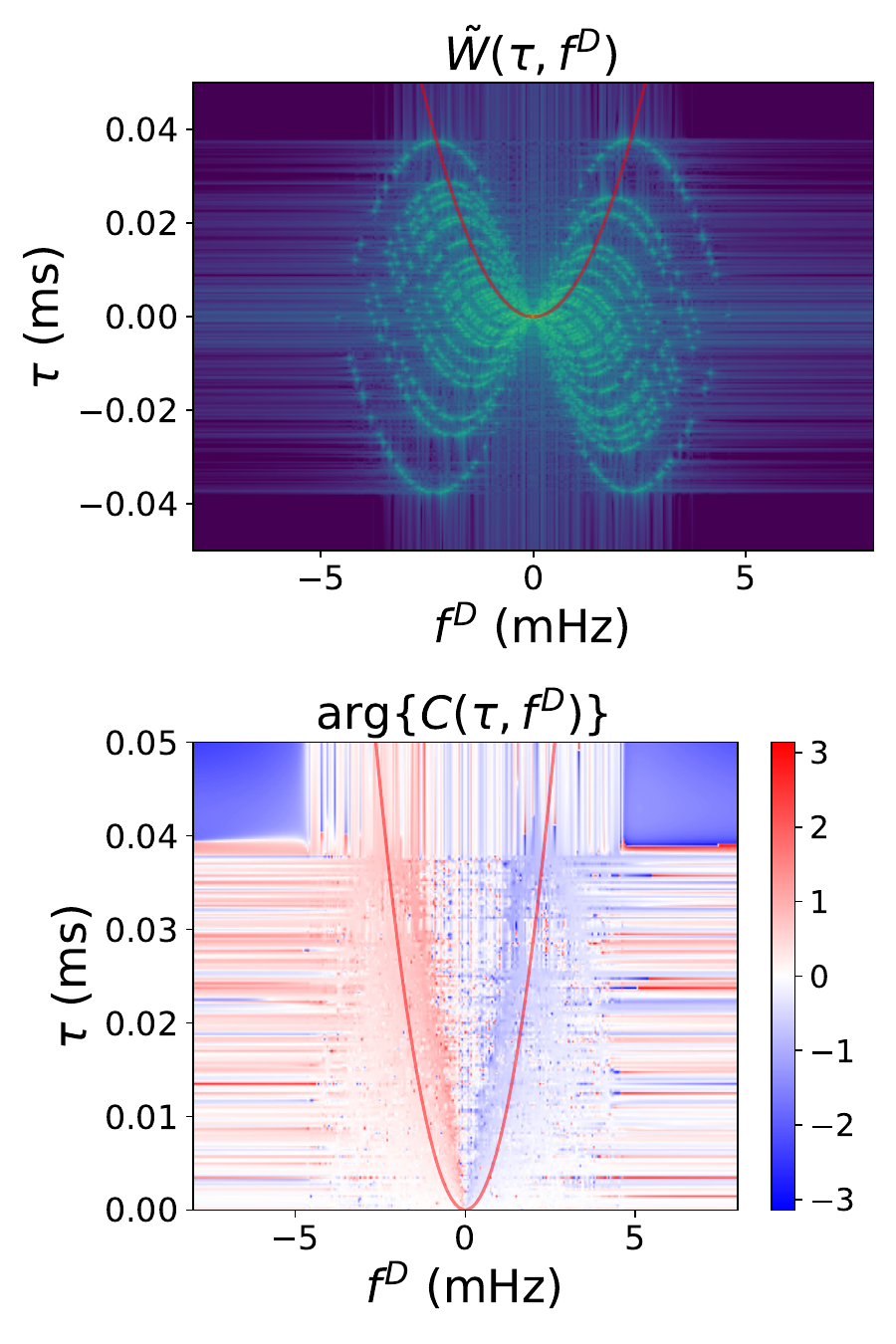}
    \caption{(Top) The cross-secondary spectrum (Eq.~\ref{eq:Wftfv}) for a one-dimensional scattering screen at $D = 100\,{\rm pc}$ with $N_{\rm im} = 100$ images randomly distributed over $40\,{\rm mas}$, with a Gaussian brightness distribution with width $\theta_{\rm sc} = 20\,{\rm mas}$ and $\lambda_0 = 75\,{\rm cm}$. (Bottom) The symmetric part of the cross-secondary spectrum phase, computed via Eq.~\ref{eq:symmetric-phase}.}
    \label{fig:cross_ss}
\end{figure}

Fig.~\ref{fig:cross_ss} shows the cross-secondary spectrum for an anisotropic scattering screen (top) and the anti-symmetric phase, $C(\tau, f^D)$, constructed via Eq.~\ref{eq:symmetric-phase} (bottom). Since the cross-secondary spectrum is the convolution of the conjugate dynamic wavefield at $-{\bm l}/2$ with the conjugate dynamic wavefield at ${\bm l}/2$, it is essentially the convolution of a parabolic arc with itself, leading to the inverted arclet structure. When the scattering is highly anisotropic, each point in the cross-secondary spectrum can be uniquely identified with a single pair of images. The points along the primary parabola (shown by the red line) are the pairs of points such that one of the points is $\theta_j = 0$. Thus, by measuring $C(\tau, f^D)$ along the central parabola, one directly measures ${\bm l} \cdot {\bm \theta}_k$ (modulo $\lambda_0$). By performing this measurement with multiple baseline pairs, one can fully infer ${\bm \theta}_k$. However, this crucially requires that the scattering is sufficiently anisotropic so that every pair of images, $(\theta_j, \theta_k)$, has a unique $(\tau_{jk}, f^D_{jk})$, and that $\tau_{jk}$ is close enough to quadratic in $f^D_{jk}$ such that the central parabola can be identified. This technique has been applied to achieve sub-${\rm mas}$ resolution on the angular positions of the scattered images of a scintillating pulsar with $\sim 5000\,{\rm km}$ baselines at $\lambda_0 = 95\,{\rm cm}$ \citep{Brisken2010, Simard2019scint}. The resolution is not set by the baselines in the array, but rather by the effective baseline formed by the Earth's effective trajectory through the spatial wavefield: $\delta \theta \sim \lambda_0 / V_{\rm eff} t_{\rm obs}$. The terrestrial baselines simply serve to break the degeneracy between the effective velocity and the angular positions of the images.



\subsection{\label{sec:FRB}FRB Scintillometry}

So far, we have argued that the fundamental underlying observable of scintillometry is the instantaneous spatial wavefield, from which the scattered image locations and effective distance, $D$, can be directly measured. We have shown that, under certain assumptions, the spatial wavefield can be replaced by the dynamic wavefield: i.e., if the interference pattern is stable over a period of time, then one can leverage the motion of the Earth, source, and screen to sample the spatial wavefield over longer baselines than what might typically be possible with terrestrial baselines. The disadvantage of this approach is that one must average over the uncertain pulse-to-pulse amplitude and phase variations of the source, erasing the crucial phase information of the wavefield. Moreover, replacing a 2D sampling of the spatial wavefield with a sampling in time over a 1D trajectory through space collapses the full, 3D $(\tau, {\bm k})$-space to a 2D $(\tau, f^D)$-space. The result is that, in general, there is not a one-to-one mapping between $(\tau, f^D)$-space and the image positions on the sky, unless the scattering is highly anisotropic. Moreover, since the Doppler shift, $f^D$, is scaled by an effective velocity, $V_{\rm eff, \parallel}$, the image positions are degenerate with this \textit{a priori} unknown parameter. Nevertheless, some of these issues can be ameliorated in some cases. For example, as shown in Section~\ref{sec:multibaseline-psr-scint}, measuring the cross-dynamic spectrum between multiple terrestrial baselines enables one to break the degeneracy between ${\bm V}_{\rm eff}$ and ${\bm \theta}_j$. Phase retrieval techniques like $\theta$-$\theta$ and cyclic spectroscopy can recover the phase and infer the dynamic wavefield from intensity data alone. These techniques have been used to reconstruct the full scattering geometry of several pulsar scintillation systems. The most well-studied example involves PSR B0834+06. \citet{Brisken2010} and \citet{Simard2019scint} use the cross-secondary spectrum technique to directly infer the image locations (and confirm the anisotropic nature of the scattering). \citet{Liu2016lensgeometry} use VLBI astrometry of the pulsar to determine its distance and velocity, thereby enabling a direct inference of the effective screen velocity and distance. \citet{HRZhu2023} use the $\theta$-$\theta$ phase-retrieval technique to infer the dynamic wavefield and fully reconstruct the lensing geometry. Nevertheless, apart from such exceptional cases, it remains challenging to measure the image locations via the dynamic wavefield, unless the scattering is highly anisotropic. 

In principle, any technique that can be applied to pulsar scintillation can be applied to FRB scintillation, provided that the FRB is a repeater (with the exception of cyclic spectroscopy, which relies on the periodic nature of pulsar signals). The primary challenge with FRBs is one of signal to noise. Nevertheless, highly sensitive telescopes such as FAST have begun to reveal the same scintillation structures (such as parabolic arcs in the secondary spectra) of scintillating FRBs \citep{Wu2024FAST}. One advantage FRBs have to pulsars in relation to scintillometry is that FRBs will typically be in the regime where either the Earth motion or the source motion dominates the effective velocity. So far, observations suggest that FRB scintillation typically arises due to scattering either in the Milky Way ISM or the host ISM. Thus, due to the large distances to FRBs, if the scattering screen is in the host ISM, then $D_l \gg D_{ls}$, and the source velocity dominates. If, however, the scattering screen is in the Milky Way, then $D_{ls} \gg D_l$ and $D_{ls} \sim D_s$, reducing Eq.~\ref{eq:Veff} to ${\bm V}_{\rm eff} = {\bm V}_\oplus - {\bm V}_l$. In this case, the primary source of uncertainty in the effective velocity (the source velocity) is removed. Likewise, when the scattering screen is in the Milky Way, the \textit{effective} distance $D$ reduces to the distance to the scattering screen, $D_l$. The source distance and velocity become irrelevant. However, measuring the dynamic wavefield alone will be insufficient to tell whether the scintillation is occurring in the Milky Way or the host ISM. Multiple simultaneous observations with terrestrial baselines will be needed to measure the angular extent of the scattering disc. 
 
Now, while the techniques of pulsar scintillometry may be applicable to scintillating, repeating FRBs, the vast majority of FRBs appear to be one-off events. As a result, one must rely on directly sampling the instantaneous spatial wavefield in order to constrain the scattering geometry. As we have discussed, in principle the spatial wavefield contains all of the possible information. However, the largest terrestrial baselines can only provide angular resolutions of $\sim 10 \,{\rm mas}$. In general, this will be insufficient to resolve individual scattered images. Nevertheless, it will be possible to place meaningful constraints on the size of the scattering disc, which can potentially alleviate ordering degeneracies in two-screen scattering measurements. Namely, scattering discs in the Milky Way will be resolved, whereas scattering discs in the host will not.

\section{Single-burst Scintillometry}
\label{sec:sims}

In the previous sections, we assumed that the instantaneous spatial wavefield takes the idealized form of Eq.~\ref{eq:ISW_sum}. This form is obtained when the intrinsic burst is a delta function; in other words, Eq.~\ref{eq:ISW_sum} is the Fourier transform of the impulse response function (IRF) \citep{T:2025pew}. Alternatively, when the bandwidth of the burst is large relative to the scintillation bandwidth\footnote{The scintillation bandwidth is defined observationally as the characteristic scale of the scintillation induced intensity modulations in the frequency domain. It is inversely related to the scattering timescale. Thus, the condition that the scintillation bandwidth be large relative to the scintillation bandwidth is equivalent to the condition that the burst duration is \textit{small} relative to the scattering times, so that the bursts are effecively delta functions.}, Eq.~\ref{eq:ISW_sum} holds. An opposite extreme is obtained when the intrinsic burst is essentially a white noise process. This occurs when both the intrinsic burst width is long relative to the scintillation timescale \textit{and} the time resolution of the detector is large relative to the period of the signal (or, equivalently, the bandwidth of the detector response is small relative to the signal bandwidth). This can occur, for example, when raw GHz signals are converted to lower-frequency baseband data.  In this case, Eq.~\ref{eq:ISW_sum} is simply modified by multiplying each frequency by a random phase $\phi(\nu)$. Doing so complicates the form of the conjugate spatial wavefield. Since $\phi(\nu)$ does not depend on the observer position, the conjugate spatial wavefield still forms delta functions in the ${\bm k}$-plane, allowing for a direct measurement of the angular position of the images. However, the time delay can no longer be measured by Fourier transforming the spatial wavefield along the frequency axis. Methods which infer the phase imparted by the scattering screen from the intensity data alone are unaffected by this change. Moreover, the cross-dynamic spectrum described in Section~\ref{sec:multibaseline-psr-scint} is unaffected since only terms that depend on either the observer position ${\bm b}$ or the image $j$ survive the conjugate multiplication in Eq.~\ref{eq:cross-secondary-spectrum}. However, as we have discussed, these techniques require observing many repeat bursts or assuming that the scattering screen is highly anisotropic (or both). Most FRB sources, however, have not been seen to repeat. For single bursts, the effects of finite intrinsic burst widths may complicate the inference of the conjugate spatial wavefield directly from the instantaneous spatial wavefield. 

\begin{figure}
    \centering
    \includegraphics{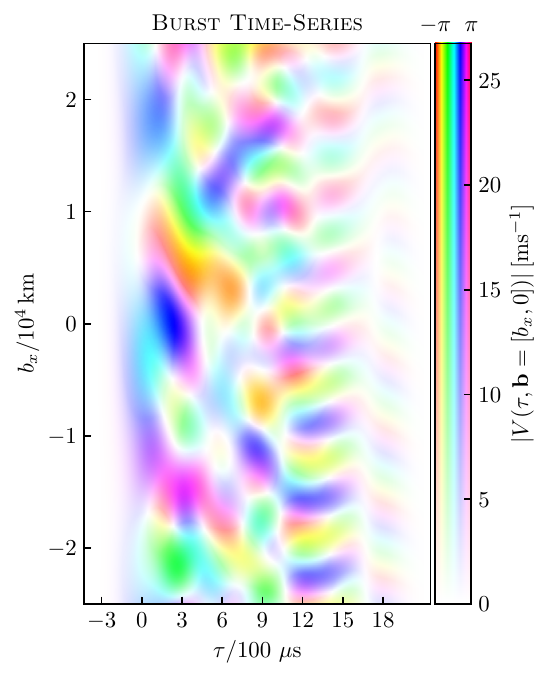}
    \includegraphics{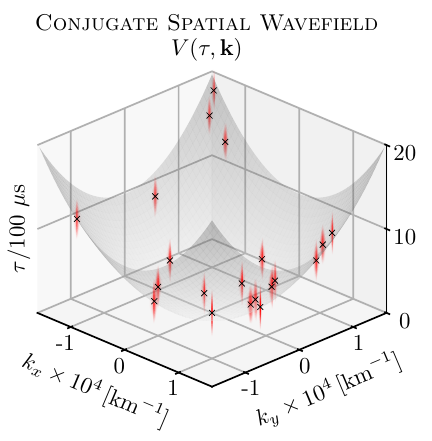}
    \caption{(Top) The burst time series, $V(\tau, {\bm b})$, of a single FRB scattered into 100 images randomly placed on an isotropic scattering disc of angular size $0.015''$ at an effective distance $D=3\,{\rm kpc}$. The colour shows the phase of the electric field and the opacity represents the amplitude (see Appendix~\ref{app:gaussian} for more details on how this is calculated). For visualization purposes, we only plot the burst time series for a single slice in baselines, ${\bm b} = (b_x,0)$.   (Bottom) The conjugate spatial wavefield computed from the burst shown in the top panel. The instantaneous spatial wavefield, $V(\nu, {\bm b})$, is the spectrum obtained from the single burst time series for fixed baseline, ${\bm b}$. Then, the 3D Fourier transform is computed to obtain the conjugate spatial wavefield (shown in red). The opacity of the conjugated spatial wavefield in each voxel is set by the ratio $|V(\tau,\vk)|/{\sf max}(|V(\tau,\vk)|)$ and voxels with opacity $<0.05$ are not plotted. One can see that the power is centred at the predicted positions of the scattered images (black crosses) in $(\tau, \vk)$-space, which form a paraboloid specified by  \Eq{tau_v_sigma}.
    } 
    \label{fig:single_burst_conjugated}
\end{figure}

In general, FRBs are scintillated in neither extreme (the observed signal is not the IRF, nor are the phases between frequency channels completely randomized). A typical intrinsic burst duration for an FRB is a millisecond, whereas typical scintillation timescales may be $\sim 0.1\,{\rm ms}$. Thus, to assess the possibility of inferring the conjugate spatial wavefield from one-off FRBs, one must model the effects of the finite, intrinsic burst profile. In Appendix~\ref{app:gaussian}, we describe a formalism for computing the observed burst time series for general intrinsic burst profiles and detector responses (and obtain an analytic form for the observed instantaneous spatial wavefield for a Gaussian intrinsic burst profile and detector response). The burst time series, $V(\tau, {\bm b})$, is the time series of the observed electric field at different baselines; its spectrum is the instantaneous spatial wavefield from which the conjugate spatial wavefield can be computed. Since the burst time series, $V(\tau, {\bm b})$, and the instantaneous spatial wavefield, $V(\nu, {\rm b})$, are related via Fourier/inverse Fourier transform along the time/frequency axis, the conjugate spatial wavefield $\tilde{V}(\tau,{\rm b})$ is simply obtained by Fourier transforming the burst time series along the two spatial axes. Fig.~\ref{fig:single_burst_conjugated} shows the burst time series and resultant conjugate spatial wavefield for a single, finite, Gaussian burst with scattering parameters defined in Appendix~\ref{app:gaussian}. The conjugate spatial wavefield was computed by sampling the burst spectrum for a uniform grid of observer positions, ${\bm b}$. A realistic configuration of baselines would be sparse and non-uniform; however, this calculation demonstrates that the parabolic structure in $(\tau, {\bm k})$-space can still be discerned even when finite burst effects are taken into account. 

\begin{figure}
    \centering
    \includegraphics{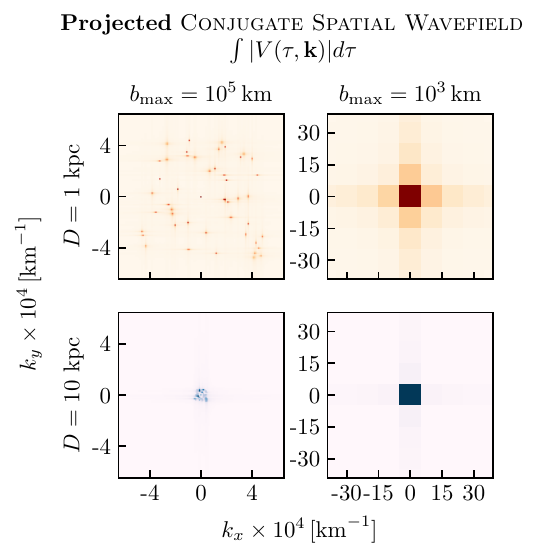}
    \caption{(Top row) The conjugate spatial wavefield projected onto the 2D ${\bm k}$-space of a single Gaussian burst with temporal width $100\,{\rm \mu s}$, scattered by a screen at $D = 1\,{\rm kpc}$ with a scattering disc of width $L = 100\,{\rm au}$, corresponding to an angular broadening of $0.1''$. The left panel shows the conjugate spatial wavefield inferred from simultaneous observations across a uniform grid of baselines out to $b_{\rm max} = 10^5\,{\rm km}$, while the right panel shows the same for a maximum baseline of $b_{\rm max} = 10^3\,{\rm km}$. (Bottom row) The same set up as the top row, except the scattering screen is placed at $D = 10\,{\rm kpc}$, corresponding to an angular broadening of $0.01''$}
    \label{fig:k-space}
\end{figure}

As we have already mentioned, finite burst effects only affect the ability to infer the group delay of the scattered images from the conjugate spatial wavefield. Inferring the angular positions of the images by measuring their positions in the ${\bm k}$-plane is unaffected. The top row of Fig.~\ref{fig:k-space} shows the conjugate spatial wavefield projected onto the 2D $k$-space for a single Gaussian burst with a $100\,{\mu s}$ width, scattered by a screen at a distance $D = 1\,{\rm kpc}$ and a scattering disc of radius $L = 100\,{\rm au}$ (corresponding to an angular broadening size of $L/D = 0.1''$). The left panel shows the inferred conjugate spatial wavefield for a uniformly sampled grid of baselines extending out to a maximum baseline of $b_{\rm max} = 10^5 \,{\rm km}$. At such large baselines, the individual images that make up the scattering disc are possible to resolve. The right panel shows the same burst but with a more realistic maximum terrestrial baseline of $b_{\rm max} = 10^3\,{\rm km}$. This can be contrasted with the bottom row of  Fig.~\ref{fig:k-space} which shows the same set up, except the scattering screen is placed at $D = 10\,{\rm kpc}$, making the effective angular size of the scattering disc $0.01''$. While extreme baselines are needed to resolve the individual scattered images, more modest baselines of $b = 10^3\,{\rm km}$ can be used to resolve the full angular extent of the scattering disc down to $\lambda / b \sim 75\,{\rm cm} /10^3\,{\rm km} \sim 0.1''$. In general, the angular extent of FRBs scattered by the Milky Way ISM will be resolved, whereas the angular extent of scattering due to the host ISM will be well below this resolution limit. Thus, multiple simultaneous observations of a single scintillated FRB across $\sim 10^3\,{\rm km}$ baselines will be sufficient to disentangle a key degeneracy in two-screen FRB scintillation studies. Namely, it will enable associations of each screen with either the Milky Way ISM or the host ISM. 

\section{\label{sec:dispersive_delay}Dispersive delays}

In the preceding sections, we demonstrate that astrometric information about the scattered images for scintillating sources is fully contained, in principle, in the instantaneous spatial wavefield. Specifically, we show that the scattered images form delta functions in the 3D Fourier transform of the instantaneous spatial wavefield (the conjugate spatial wavefield). In this $(\tau, {\bm k})$-space, the angular positions of the images are directly proportional to the ${\bm k}$ coordinate. When the time delay is dominated by the geometric component, the scattered images form a paraboloid in $(\tau, {\bm k})$-space (Eq.~\ref{eq:tau_v_sigma}). In reality, every image acquires an additional frequency-dependent dispersive delay, $\zeta_j$. This dispersive delay arises not only from the scattering screen itself, but also each image traverses a different path between the source and observer. Differences in the total electron column along that path lead to different dispersive delays. 

The dispersion measure (DM) of a pulsar or FRB is a direct measurement of the total electron column density along the line of sight, ${\rm DM} = \int n_e dz$. As radio waves propagate through an ionized medium, the waves are dispersed, with lower frequencies arriving later than higher frequencies. The effect is a frequency dependent dispersive time delay:
\begin{equation}
    \zeta = \frac{\lambda^2 r_e {\rm DM}}{2 \pi c}.
\end{equation}
Pulsar and FRB observations are typically \textit{de}-dispersed before characterizing effects such as scintillation and scattering, i.e. the frequencies are shifted to correct for the dispersion due to the overall ${\rm DM}$. However, by measuring the differential dispersive delays associated with each image, $\zeta_j$, one can potentially probe fluctuations in ${\rm DM}$ on small transverse scales. For example, consider a Milky Way scattering screen at a distance, $D_l = 1\,{\rm kpc}$, which produces images out to a maximum angle of $\theta \sim 100\,{\rm mas}$. The physical separation of these images is $\sim 100\,{\rm au}$. Thus, by measuring a gradient in the dispersive delay of the images across the screen, one is directly probing the variance in the ${\rm DM}$ towards the FRB for sight-lines separated by $\sim 100\,{\rm au}$. 

In Appendix~\ref{sec:DMstructure}, we compute the ${\rm DM}$ structure function for a turbulent medium with a power spectrum of 3D density fluctuations given by $P(k) = C_n^2 k^{-\beta}$. The ${\rm DM}$ structure function is the variance of the \textit{projected} density between sight-lines at some fixed transverse separation. We find that for sight-lines through a medium of thickness, $L$, the rms fluctuations in the electron column density is given by 
\begin{equation}
\Delta {\rm DM}(r_\perp) =\sqrt{2 A_\beta B_\beta C_n^2 L r_\perp^{\beta-2} / \pi^2}, 
\end{equation}
where $r_\perp$ is the transverse separation between the two sight-lines. The parameters $A_\beta$ and $B_\beta$ are constants that are independent of the inner and outer scales of the turbulence and are of order unity for $3 < \beta < 4$. For Kolmogorov turbulence ($\beta = 11/3$), we find 
\begin{equation}
    \Delta {\rm DM}(r_\perp) = \sqrt{0.36 C^2_n L r^{5/3}_\perp}.
    \label{eq:delta_DM}
\end{equation}
The sight-line toward any given FRB will intersect various ionized mediums, including the Milky Way and host ISM, the Milky Way and host halos, the IGM, and any intersecting halos along the line of sight. In what follows, we will estimate the magnitude of the contribution to the dispersive delays for these different mediums. Table~\ref{tab:DMstructure} summarizes the estimates we obtain.

\begin{table*}[]
    \centering
    \begin{tabular}{|c|c|c|c|}
        \hline
         Medium & $C^2_n$ (${\rm m}^{-20/3}$) & $\Delta {\rm DM}$ $({\rm pc\,cm^{-3}})$ & $\zeta$ $(\mu {\rm s})$ \\
         \hline
         MW ISM & $10^{-4}$ & $10^{-4}$ & 1 \\
         Cool CGM & $10^{-8} - 10^{-4}$ & $10^{-6} - 10^{-4}$ & $0.01 -1 \times (1 + z_{\rm halo})^{-2}$ \\
         Hot CGM & $10^{-16} - 10^{-8}$ & $10^{-10} - 10^{-6}$ & $\lesssim 0.01 \times(1+z_{\rm halo})^{-2}$ \\
         IGM & $< 10^{-12}$ & $<10^{-5} $& $\lesssim 0.1$ \\
         Host ISM & $10^{-4}$ & $10^{-4}$ & $1 \times (1 + z_{\rm FRB})^{-2}$ \\
         \hline
    \end{tabular}
    \caption{Order-of-magnitude estimates for the contribution to the differential ${\rm DM}$ and resulting dispersive delay through sight-lines ($\lambda = 75\,{\rm cm}$) separated by $r_\perp = 100\,{\rm au}$ through different ionized mediums. Values for the cool CGM are computed assuming a halo of $r_{\rm vir} \sim 200\,{\rm kpc}$ and a volume filling factor of $f \sim 10^{-2}$.}
    \label{tab:DMstructure}
\end{table*}

\subsection{\label{sec:ISM_dispersion}The Milky Way and Host ISM}

For the ionized component of the ISM, we take the amplitude of the density fluctuations to be $C_n^2 \sim 10^{-4} {\rm m}^{-20/3}$, consistent with the value inferred from temporal variations in the dispersion measure of pulsars \citep{Ocker2021}. Then, Eq.~\ref{eq:delta_DM} gives:
\begin{align}
\begin{split}
    \Delta {\rm DM}_{\rm ISM} \sim &1 \times 10^{-4} \, {\rm pc \, cm^{-3}} \, \\ \times &\left( \frac{C_n^2}{10^{-4} \, {\rm m}^{-20/3}} \right)^{1/2} \left( \frac{L}{{\rm kpc}} \right)^{1/2} \left( \frac{r_\perp}{100 \,{\rm au}} \right)^{5/6}.
\end{split}
\end{align}
In other words, typical sight-lines through the Milky Way ISM, separated by $r_\perp \sim 100\,{\rm au}$, will have dispersion measures that differ by $\sim 10^{-4}\,{\rm pc\,cm^{-3}}$. This results in a differential dispersive time delay of
\begin{equation}
    \zeta_{\rm ISM} = \frac{\lambda^2 r_e \Delta {\rm DM}}{2 \pi c} = 2.7\,{\rm \mu s}\,\left( \frac{\lambda}{75\,{\rm cm}} \right)^2 \left( \frac{\Delta {\rm DM}}{10^{-4} \,{\rm pc \, cm^{-3}}} \right).
\end{equation}
Note that this should be regarded as a rough lower bound on the differential dispersive delay. The amplitude of the inferred density power spectrum may be larger by orders of magnitude depending on the specific region of the ISM being probed \citep{Ocker2021}. In particular, sight-lines through the galactic centre may differ significantly from sight-lines at high galactic latitude. Nevertheless, our estimate of $\Delta {\rm DM}_{\rm ISM} \sim 10^{-4}\,{\rm pc\,cm^{-3}}$ is broadly consistent with the value inferred by \citet{Baker2025dmgradient}, who use phase retrieval to measure the DM gradient across the scintillation screen for PSR B0834+06.

Now, assuming that the FRB host ISM is similar to the Milky Way ISM, we also take $C_n^2 \sim 10^{-4} {\rm m}^{-20/3}$. Thus, the $\Delta {\rm DM}$ is the same as above. However, when converting this to a time delay, we need to use the wavelength in the rest frame of the host, as opposed to the observed wavelength:
\begin{equation}
    \zeta_{\rm ISM,\,host} = 2.7\,{\rm \mu s}\,(1 + z_{\rm FRB})^{-2}\,\left( \frac{\lambda}{75\,{\rm cm}} \right)^2 \left( \frac{\Delta {\rm DM}}{10^{-4}\,{\rm pc \, cm^{-3}}} \right).
\end{equation}
Thus, for high redshift FRBs, the contribution from the host ISM may be significantly suppressed.

\subsection{\label{sec:CGM_dispersion}The CGM}

In general, a sight-line towards a given FRB will be dispersed by gas in its host halo, the Milky Way halo, and any halo it happens to intersect. We refer to the gas within a galaxy halo as the circumgalactic medium (CGM). It is now well established that the CGM is multi-phase, with a cool $(T \sim 10^4\,{\rm K}$) component and a hot $(T \sim 10^6\,{\rm K})$ component \citep{tumlinson2017circumgalactic}. While the properties of these two components are poorly constrained, we will estimate the contribution to the differential dispersion measure using some nominal values for the turbulent fluctuations. 

\citet{ocker2025microphysics} argue that the cool phase of the CGM may have a $C_n^2$ anywhere from $\sim 10^{-8} {\rm m}^{-20/3}$ to $\sim 10^{-6}\,{\rm m}^{-20/3}$. We will consider the upper limit of this range as that will set the minimum sensitivity needed to detect a dispersion gradient in the CGM. Now, while the cool gas in the CGM is thought to have a high covering fraction \citep{Rigby2002, Hennawi2006, BowenCheclouch2011}---so that most sight lines within a virial radius of a galaxy will intersect some such gas---it is likely to have a very small volume filling fraction, $f \sim 10^{-3} - 10^{-2}$. The effective path length through the cool gas depends on this volume filling fraction, $L \approx f r_{\rm vir}$, where $r_{\rm vir}$ is the virial radius of the halo (assuming the gas suffuses the virial halo roughly uniformly). Thus, the contribution to the differential ${\rm DM}$ from an intersecting halo is
\begin{align}
    \Delta {\rm DM}_{\rm halo} \sim & 6 \times 10^{-5}\, {\rm pc \, cm^{-3}} \, \left( \frac{C_n^2}{10^{-5} \, {\rm m}^{-20/3}} \right)^{1/2} \\
    &\times \left( \frac{f}{10^{-2}} \right)^{1/2} \left( \frac{r_{\rm vir}}{200\,{\rm kpc}} \right)^{1/2} \left( \frac{r_\perp}{100 \,{\rm au}} \right)^{5/6},
\end{align}
which results in a differential dispersive delay of
\begin{equation}
    \zeta_{\rm halo} \sim 2\,{\rm \mu s} \, (1+ z_{\rm halo})^{-2} \left( \frac{\lambda}{75\,{\rm cm}} \right)^2 \left(\frac{\Delta{\rm DM}}{6 \times 10^{-5}\,{\rm pc\,cm^{-3}}} \right).
\end{equation}
Note that, given the highly unconstrained nature of the physical parameters of the cool gas, these values should only be taken as bench-mark estimates for a typical halo. The fluctuation amplitude, $C_n^2$, and the volume filling factor, $f$, may both vary significantly with halo mass and redshift, for example. In any case, the cool component of the CGM is likely to dominate the contribution to the differential dispersion as the hot component is expected to have $C_n^2 \sim 10^{-12} \,{\rm m^{-20/3}}$ \citep{ocker2025microphysics}. Even with a volume filling fraction of $f \sim 1$, the resulting dispersive delay is on the order of nanoseconds. 

\subsection{\label{sec:IGM_dispersion}IGM}

The amplitude of the density power spectrum of the IGM is poorly constrained, especially on the relevant scales. However, assuming that the fluctuations are smaller than the hot phase of the CGM, we can take $C_n^2 < 10^{-12}\,{\rm m^{-20/3}}$ for the IGM. Even allowing for the large path lengths through the IGM  ($L \sim {\rm Gpc}$), one obtains a differential dispersion measure of $\Delta {\rm DM} \ll 10^{-5}\,{\rm pc \,cm}^{-3}$. Thus, the IGM contribution to the differential dispersion is likely to be negligible. 

\subsection{Observability}

\citet{Baker2025dmgradient} perform the first measurement of a dispersion measure gradient across a scintillation screen for a pulsar. In doing so, they directly infer the amplitude of the DM structure function for that particular sight-line, which we find to be broadly consistent with our estimate for the Milky Way ISM. Thus, leveraging scintillometry in this way to measure differential dispersive delays between scattered images, one can probe the microphysics of turbulent ionized mediums on tiny transverse scales ($r_\perp \sim 100\,{\rm au}$). Applying this technique to pulsars will shed light on the ISM; however, due to their cosmological origins, FRB scintillometry may be used to probe the DM structure function for the cool gas phase of the CGM. The physics and morphology of the cool CGM phase is highly unconstrained and may be directly relevant to central questions of galaxy evolution and star formation \citep{tumlinson2017circumgalactic, FaucherGiguere2023}. 

We have estimated that the contribution to the differential dispersive delay from a single, low-redshift CGM halo will be at most $\sim 1\,{\rm \mu s}$ at $\lambda = 75\,{\rm cm}$, which is of comparable size to the Milky Way ISM contribution. In general, however, an FRB sight-line will intersect multiple CGM halos, including its host halo, and the Milky Way CGM. High-redshift sources ($z_s > 1$) may intersect multiple intervening halos along the line of sight. Contributions to the DM gradient across the scintillation screen add in quadrature. However, the excess dispersive delay from individual halos decreases with redshift by a factor of $(1 + z_l)^2$, so that the differential dispersion will generally be dominated by low-redshift halos. Nevertheless, we anticipate high redshift FRBs to have a systematically larger differential dispersion than their low redshift counterparts, and FRBs in general to have a systematically larger differential dispersion than pulsars. The magnitude of this difference will depend on the actual value of $C^2_n$ for the intervening halos.

Measuring the differential dispersion from the CGM on $r_\perp \sim 100\,{\rm au}$ scales will require being able to resolve $\sim 1\,{\rm \mu s}$ dispersive delays. Specifically, one will need to be able to measure the \textit{geometric} delay of the scattered images to within a microsecond. Since the geometric delay is quadratically related to the angular position of a scattered image via $\tau= D\theta^2/2c$, in order to infer the geometric delay to a precision of $\delta \tau$, one must be able to measure the image location to a precision of 
\begin{equation}
    \delta \theta = \frac{c \delta \tau}{2 r_\perp},
\end{equation}
where we have replaced $r_\perp = D \theta$ (i.e. the transverse size of the scintillation screen). The achievable angular resolution on the images is set by the effective baseline, $b$, that one is able to sample the spatial wavefield over via $\delta \theta = \lambda / b$. Thus, to achieve a precision on the time delay, $\delta \tau$, one requires a minimum baseline of
\begin{equation}
    b_{\rm min} = \frac{2 r_\perp \lambda}{c \delta \tau}.
\end{equation}
For $\delta \tau \sim 1\,{\rm \mu s}$ and $r_\perp \sim 100\,{\rm au}$, one must be able to resolve the scattered images to a precision of $\delta \theta \sim 2\,{\rm \mu as}$, which corresponds to a baseline of $b_{\rm min} \sim 10^8 \,{\rm km}$ at $\lambda = 75\,{\rm cm}$. Such baselines are not possible to achieve with simultaneous terrestrial observations. However, repeating FRBs observed over a few months may be able to achieve similar effective baselines. The stability of the scintillation pattern over this timescale, however, is less certain. The proposed measurement may become more achievable at lower frequencies, since as we have seen, the dispersive delay scales as $\Delta \sim \lambda^2 r^{5/6}_\perp$. The size of the scintillation screen itself scales with wavelength as $r_\perp \sim \lambda^2$ (since the refractive bending angle of a plasma lens increases as $\lambda^2$). Thus, the required minimum baseline scales as $b_{\rm min} \sim \lambda^{-2/3}$. In other words, although the achievable angular resolution for a given baseline worsens with increasing $\lambda$, the size of the dispersive delay increases faster, so that shorter baselines are needed at longer wavelengths.

\section{\label{sec:concl}Conclusion}

In this paper, we have introduced the instantaneous spatial wavefield as the underlying observable of scintillometry that contains all of the pertinent astrometric information. We have related the instantaneous spatial wavefield to more familiar quantities in the pulsar scintillometry literature. We claim that the underlying assumptions of different scintillometry techniques are clarified by viewing these quantities as samplings of the conjugate spatial wavefield. This enables us to further consider prospects for applying pulsar scintillometry techniques to the emerging field of FRB scintillometry. In particular, we argue that full lensing geometry reconstructions (i.e. full inferences of the scattered image locations and time delays) will be achievable only with high signal-to-noise, repeating FRBs. However, multiple observations across terrestrial baselines will at least be sufficient to instantaneously resolve the extent of the scattering discs for one-off FRBs. This will break a key degeneracy in two-screen FRB scintillation constraints (i.e. which scattering screen is closer to the observer and which is closer to the source). In the cases where full lensing geometries can be reconstructed, differential dispersive delays between the scattered images may be measured, enabling measurements of DM gradients for sight-lines separated by transverse distances of $\sim 100\,{\rm au}$. In other words, one may potentially measure the DM structure function on very small transverse scales. We estimate the potential contributions to this structure function and show that the contribution from intervening CGM halos may be as large as the Milky Way ISM contribution. This would open up a new means of constraining the physics of the turbulent, cool phase of the CGM at tiny scales.

\section*{Acknowledgments}
We would like to thank Tim Sprenger, Kenzie Nimmo, and Diego Montalvo for helpful discussion.
This research made use of computational resources at SLAC National Accelerator Laboratory, a U.S. Department of Energy Office of Science laboratory, and the Sherlock cluster at the Stanford Research Computing Center (SRCC). 
We would like to thank Stanford University, SRCC, and SLAC for providing computational resources that contributed to these research results.
D.S. is additionally supported by the National Science Foundation Graduate Research Fellowship under Grant No. DGE-2146755.

\appendix

\section{DM Structure Function}
\label{sec:DMstructure}

Consider a free electron density field $n_e({\bm r})$. The two-point correlation function and corresponding power spectrum is
\begin{align}
    \xi({\bm r}) &= \langle n_e({\bm x}) n_e({\bm x}+{\bm r})\rangle, \\
    P({\bm k}) &= \int \xi({\bm r})e^{i {\bm k} \cdot {\bm r}} d^3 r.
\end{align}
We will assume that the density field is isotropic such that $\xi({\bm r}) = \xi(r)$ and $P({\bm k} ) = k$. We will also assume the density fluctuations obey some power law, $P(k) = C^2_n k^{-\beta}$, between an inner scale $l_i$ and outer scale $l_o$. For Kolmogorov density fluctuations $\beta = 11/3$. 

The observed electron column density, which is probed by the dispersion measure, is given by
\begin{equation}
    N_e({\bm r}_\perp) = \int n_e({\bm r}_\perp, z) dz,
\end{equation}
where $z$ is the coordinate along the line of sight and ${\bm r}_\perp$ is the 2D vector transverse to the line of sight. We would like to compute the column density structure function:
\begin{align}
    \nonumber
    D_N(r_\perp) &= \langle \left[ N({\bm x}_\perp) - N({\bm x}_\perp + {\bm r}_\perp) \right]^2\rangle, \\
    &= 2 \left[ \xi_N(0) - \xi_N(r_\perp)\right],
\end{align}
where we define the projected two-point correlation to be $\xi_N(r_\perp) = \langle N_e({\bm x}_\perp+{\bm r}_\perp) N_e({\bm x}_\perp) \rangle$. Again, due to isotropy, $\xi_N({\bm r}_\perp) = \xi_N(r_\perp)$. 

Let us consider a slab of free electrons of depth $L$. Using the definition of $N_e$, one obtains
\begin{align}
    \nonumber
    \xi_N(r_\perp) &= \int_{-L}^L \int_{-L}^L dz dz' \xi({\bm r}_\perp, z-z'), \\
    &= \int_{-L}^L dZ (L - Z) \xi\left(\sqrt{r_\perp^2 + Z^2}\right).
\end{align}
The latter follows from the change of variables $z \to Z = z-z'$. Thus, the structure function is
\begin{align}
    D_N(r_\perp) &= 2 \int dZ (L-Z) \left[ \xi(Z) - \xi\left(\sqrt{r^2_\perp + Z^2}\right)\right]. 
\end{align}
Note that the term in the brackets can be trivially rewritten as
\begin{equation}
    \left[ \left(\xi(0) - \xi(\sqrt{r^2_\perp + Z^2})\right) - \left(\xi(0) - \xi(Z)\right)\right].
\end{equation}
This can be simplified by noting that
\begin{align}
    \nonumber
    \xi(0) - \xi(r) &= \frac{1}{2 \pi^2} \int dk k^2 P(k) \left( 1 - \frac{\sin(kr)}{kr}\right), \\
    \nonumber
    &= \frac{C^2_n}{2 \pi^2} \int dk k^{2-\beta} \left( 1 - \frac{\sin(kr)}{kr}\right), \\
    &= \frac{C^2_n A_\beta}{2 \pi^2} r^{\beta -3},
\end{align}
where the last line follows from the substitution $k 
\to u = kr$, and 
\begin{align}
    A_\beta \equiv  \int_0^\infty  du u^{2 - \beta} \left( 1 - \frac{\sin u}{u} \right) = -\Gamma(2-\beta) \sin\left(\frac{\pi \beta}{2}\right).
\end{align}
This integral is convergent for $3 < \beta < 5$. For Kolmogorov turbulence, we obtain $A_{11/3} \approx 1.2$. 
It now follows that
\begin{equation}
D_N(r_\perp) \approx \frac{2 A_\beta B_\beta}{\pi^2} C_n^2 L r^{\beta-2},
\end{equation}
where we have assumed $l_o\ll L,r_\perp$ and have defined
\begin{align}
    \nonumber
    B_\beta &\equiv \int_0^\infty dv \left[(1+v^2)^{(\beta-3)/2} - (v^2)^{(\beta-3)/2} \right] \\
    &= \frac{\sqrt{\pi}}{2}\frac{\Gamma(1 - \beta/2)}{\Gamma(3/2 - \beta/2)}.
\end{align}
This integral is convergent whenever $\beta < 4$, since the integrand is asymptotically $\sim v^{\beta - 5}$ as $v \to \infty$. Thus, we have a simple expression for the electron column density structure function for transverse displacements less than the outer scale and a power-law spectrum of density fluctuations with $3 < \beta < 4$:
\begin{equation}
    D_N(r_\perp) = \frac{2^{1-\beta}}{\pi^2} \Gamma\left(1 - \frac{\beta}{2}\right)^2 \sin\left(\frac{\pi}{2}(4 - \beta)\right) C_n^2 L r_\perp^{\beta - 2}.
\end{equation}
Thus, the typical rms fluctuations of the electron column density through a turbulent slab of thickness, $L$, at transverse separations, $r_\perp$, is given by
\begin{equation}
    k_{N_e} (r_\perp)  = \sqrt{0.36 C^2_n L r_\perp^{5/3}},
\end{equation}
for a Kolmogorov turbulent medium ($\beta = 11/3$).

\begin{figure}[t]
    \centering
    \includegraphics{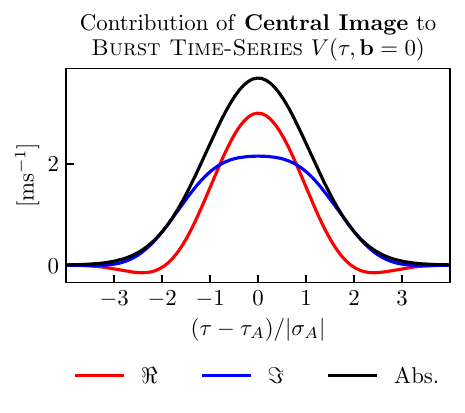}
    \caption{
    The magnitude (black) of the central $\vtheta=(0,0)$ image contribution to the burst time-series (\Eq{appEtb}) along with its real (red) and imaginary (blue) component for a detector at baseline $\vb = 0$ assuming the intrinsic and burst and detector responses are Gaussian.
    The resulting burst time series with contribution from many images including the central image shown here is visualized for $\vb=(0,0)$ in \Fig{all_images} and for $\vb=(b_x,0)$ in the top panel of \Fig{single_burst_conjugated} in the main text.
    }
    \label{fig:single_image}
\end{figure}

\begin{figure}[t]
    \centering
    \includegraphics{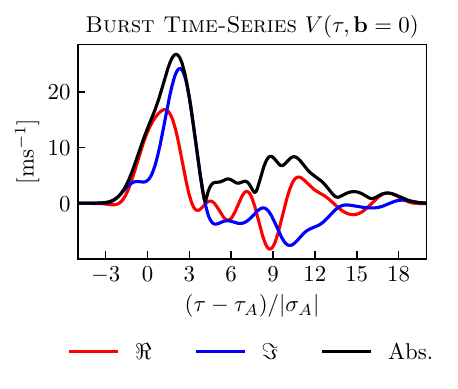}
    \caption{
    The magnitude (black) of the burst time-series (\Eq{appEtb}) along with its real (red) and imaginary (blue) component from one central ($\vtheta=(0,0)$, see also \Fig{single_image}) and 99 randomly chosen points on the scattering screen for a detector at baseline $\vb = 0$ assuming the intrinsic and burst and detector responses are Gaussian.
    We visualize more generally the burst time series as a function of $b_x$ in $\vb=(b_x,0)$ in the top panel of Fig.~\ref{fig:single_burst_conjugated} in the main text.
    }
    \label{fig:all_images}
\end{figure}

\section{Wavefield of a single Gaussian burst}
\label{app:gaussian}

{Here, we compute the spatial wavefield for a single, finite burst. We will assume the FRB has an intrinsic Gaussian profile, $\psi(t)$, and some central frequency $\nu_\psi$: }
\begin{equation}
   \psi(t) \sim e^{-t^2 / 2 k_\psi^2}  e^{-i2\pi\nu_\psi t} \Rightarrow \wt\psi(\nu) = e^{-(2\pi(\nu-\nu_\psi))^2k_\psi^2 / 2}.
\end{equation}
{We assume that the pulse encounters a scattering screen with some effective distance, $D$, which forms multiple images, $\theta_j$. Each of these images will arrive at the observer with a geometric time delay:}
\begin{equation}
\tau^{\rm geom.}_j(\vb) =  \frac{D}{2c}\vtheta_j^2 - \frac{\vb\cdot\vtheta_j}{c}.
\end{equation}
{We also assume that the screen imparts a magnification $\mu_j$ to the image formed at the angular location $\theta_j$, which we take to have a Gaussian profile:}
\begin{equation}
    \mu_j^{1/2} \sim e^{-\theta_j^2 / 2 k_\theta^2},
\end{equation}
{where $k_\theta$ characterizes the angular size of the screen\footnote{In principle, the magnification will also depend on frequency but we neglect this for simplicity.}.}
We introduce a length scale for the size of the screen $L$ such that $Dk_\theta\sim L$.

{As the pulse propagates through an ionized medium, it will also accumulate a frequency dependent phase $\varphi$, which depends on the dispersion measure along the path from source to observer. Since each image travels along a different path, the accumulated phase also depends on the image, $\varphi_j(\nu)$.} We define the temporal delay kernel $\D_j(t)$ due to this DM as 
\begin{equation}
   \D_j(t) = \int d\nu\, e^{i(\varphi_j(\nu) - 2\pi\nu t)}\Leftrightarrow\wt\D_j(\nu) = e^{i\varphi_j(\nu)}.
\end{equation}

{We would like to be able to model the signal a realistic detector would measure for such a scintillated burst. 
We will assume the burst is measured in a frequency channel of our detector with spectral response $\wt\chi(\nu)$. The voltage measured by the detector at a position $\vb$ is then}
\begin{equation}
    V(\vb, \nu) = \sum_j \mu_j^{1/2} \wt\chi(\nu)\wt \D_j(\nu) \wt \psi(\nu) e^{i2\pi\nu \tau^{\rm geom.}_{j}(\vb)}.
\end{equation}
{This is equivalent to the instantaneous spatial wavefield that we defined in Eq.~\ref{eq:ISW_sum}, except now we have taken into account the finite temporal extent of the intrinsic burst and the detector response. 
Eq.~\ref{eq:ISW_sum} is recovered when the intrinsic burst and the detector response are both delta functions, so that their spectra are uniform.}

We can take the inverse Fourier transform of this spatial wavefield along the $\nu$ axis to obtain the \textbf{burst time series}: $V(\tau,\vb)$. 
Each individual image's contribution to this time series is a convolution (denoted by $\star$) of the detector response, dispersion, and intrinsic pulse delayed by the appropriate geometric contribution $\tau_j^{\rm geom.}$:
\begin{equation}
    V(\tau, \vb) = \sum_j \mu_j^{1/2} [\chi\star \D_j \star \psi](\tau-\tau^{\rm geom.}_{j}(\vb)).
    \label{eq:appEtb}
\end{equation}

Analytical progress in evaluating this equation can be made with some approximations. Consider a single frequency channel in our detector with a Gaussian spectral response function centred around a frequency $\nu_A$:
\begin{equation}
    \wt \chi_A(\nu) = \wt \Delta(\nu-\nu_A)= {e^{-(\nu-\nu_A)^2 / 2 k_{\wt\Delta}^2}}.
\end{equation}
We approximate the accumulated phase due to the dispersive delay, $\varphi_j$, as quadratic around this central frequency $\nu_A$:
\begin{equation}
    \varphi_{Aj}(\nu) \approx \varphi_j(\nu_A) - 2\pi \zeta_{Aj}(\nu - \nu_A) + \pi{\mathfrak D_{Aj}}  (\nu - \nu_A)^2,
\end{equation}
where $\zeta_{Aj}$ is the group (stationary-phase) time delay 
\begin{equation}
\zeta_{j}(\nu) = -\frac 1 {2\pi} \frac{d\varphi_j(\nu) }{ d\nu},
\end{equation}
evaluated at $\nu=\nu_A$, and $\mathfrak D_{Aj}=-d\zeta_j/d\nu\lvert_{\nu=\nu_A}$ parametrizes the pulse broadening due to dispersion and is related to the usual dispersion measure through\citep{Hirata:2013era}:
\begin{equation}
    \frac{\mathfrak D_{Aj}}{\rm s\ GHz^{-1}} = -8 \left(\frac {\rm DM}{10^3\ {\rm pc\ cm}^{-3}}\right) \left(\frac{\nu_A}{\rm GHz} \right)^{-3}.
    \label{eq:hirata_D}
\end{equation}
This allows us to evaluate the burst time series in \Eq{appEtb}, revealing that each image's contribution to $V(\tau,\vb)$ has the form of a complex Gaussian temporally delayed by
\begin{equation}
   \tau_{Aj} = \tau_j^{\rm geom.} - \zeta_{Aj}+ {\mathfrak D_{Aj}(\nu_\psi-\nu_A)},
   \label{eq:tauj}
\end{equation}
with complex width 
\begin{equation}
k_{Aj}^2 = k_\psi^2 +  \frac {1}{4\pi^2k_{\wt \Delta}^2}  + \frac{\mathfrak D_{Aj}}{2\pi i}.
\end{equation}
We denote these quantities evaluated for an image at $\vtheta=0$ as $\tau_A$ and $k_A$.
We visualize our analytical form of the burst time series for
\begin{enumerate}
    \item a central image ($\vtheta=0$) at a detector at $\vb=0$ in \Fig{single_image},
    \item a central image ($\vtheta=0$) in addition to 99 randomly positioned images on the screen uniformly chosen from $[-L/2, L/2]$ at a detector at $\vb=0$ in \Fig{all_images}, and
    \item a central image ($\vtheta=0$) in addition to 99 randomly positioned images on the screen as a function of $b_x$ where $\vb=(b_x,0)$ in the top panel of \Fig{single_burst_conjugated}.
\end{enumerate}
For the figures in this Appendix we adopt 
\begin{align*}
    \nu_\psi &= \nu_A = 600\ {\rm MHz},\\
    D &= 3\ {\rm kpc},\\
    L &= 100\ {\rm AU},\\
    k_\psi &= 100\ \mu{\rm s},\\
    k_\theta &= 0.015\arcsec,\\
    k_{\wt\Delta} &= 1\ {\rm MHz}\\
    {\rm DM} &= 10^3\ {\rm pc/cm}^3.
\end{align*}

Analytically evaluating the conjugate spatial wavefield $\mathcal F_\vb[V_A(\tau,\vb)]$ is difficult; however, the time-series $V_A(\tau,b)$ (Eq.~\ref{eq:appEtb}) can be straightforwardly sampled and the resulting Fourier transform computed numerically.
We conservatively choose the temporal sampling rate to be $10/k_{\wt \Delta}$ and assume a uniform sampling of $\vb$.
The conjugate spatial wavefield can then be computed by applying FFT along the $\vb$ axes---however, we note that for more realistic baseline distributions non-uniform FFT's \citep{1989ApJ...338..277P} would be required. 
Doing this in \Fig{single_burst_conjugated} confirms that scintillated images form a paraboloid specified by \Eq{tau_v_sigma} in $(\tau, \vk)$-space even when the conjugated spatial wavefield is computed assuming the intrinsic and burst and detector responses are Gaussian as opposed to Dirac delta's as was assumed when deriving \Eq{tau_v_sigma}.

\bibliography{main}{}
\bibliographystyle{aasjournalv7}

\end{document}